\documentclass[10pt,preprint2]{aastex}
\usepackage{psfig}
\usepackage{epsf}

\shorttitle{The All-Sky X-ray Cluster Dipole}
\shortauthors{Kocevski, Mullis \& Ebeling}

\received{2004 January 17}
\accepted{2004 February 23}
\begin{document}

\title{The Dipole Anisotropy of the First All-Sky X-ray Cluster Sample}
\author{Dale D. Kocevski, Christopher R. Mullis\altaffilmark{1}, and Harald Ebeling}

\affil{Institute for Astronomy, University of Hawaii, 2680 Woodlawn Dr., Honolulu, HI 96822}
\email{kocevski@ifa.hawaii.edu; cmullis@eso.org; ebeling@ifa.hawaii.edu}

\altaffiltext{1}{ESO Headquarters, Karl-Schwarzschild-Strasse 2, Garching D-85748, Germany.}

\begin{abstract}

We combine the recently published CIZA galaxy cluster catalogue with the XBACs cluster sample to produce the first all-sky catalogue of X-ray clusters in order to examine the origins of the Local Group's peculiar velocity without the use of reconstruction methods to fill the traditional Zone of Avoidance.  The advantages of this approach are (i) X-ray emitting clusters tend to trace the deepest potential wells and therefore have the greatest effect on the dynamics of the Local Group and (ii) our all-sky sample provides data for nearly a quarter of the sky that is largely incomplete in optical cluster catalogues.  We find that the direction of the Local Group's peculiar velocity is well aligned with the CMB as early as the Great Attractor region 40 $h^{-1}$ Mpc away, but that the amplitude of its dipole motion is largely set between 140 and 160 $h^{-1}$ Mpc.  Unlike previous studies using galaxy samples, we find that without Virgo included, roughly $\sim70\%$ of our dipole signal comes from mass concentrations at large distances ($>60$ $h^{-1}$ Mpc) and does not flatten, indicating isotropy in the cluster distribution, until at least 160 $h^{-1}$ Mpc.  We also present a detailed discussion of our dipole profile, linking observed features to the structures and superclusters that produce them.  We find that most of the dipole signal can be attributed to the Shapley supercluster centered at about 150 $h^{-1}$ Mpc and a handful of very massive individual clusters, some of which are newly discovered and lie well in the Zone of Avoidance.

\end{abstract}

\keywords{cosmic microwave background --- galaxies: clusters: general ---  large-scale structure of universe --- X-rays: galaxies: clusters}

\section{Introduction}

Since the detection of a dipole anisotropy in the Cosmic Microwave Background (CMB, Kogut 1993), many question have been raised regarding the origin of the Local Group (LG) motion which is thought to give rise to such a signal.  Specifically at question have been the nature of the cosmic objects or structures most directly responsible for inducing the LG's peculiar velocity and, furthermore, the distance out to which inhomogeneities in the distribution of these objects continue to affect the LG's dynamics.  Interest in these questions is motivated by the cosmological implications they carry; for example, if the motion of the Milky Way (MW) is induced entirely nearby, then the current density of the universe would need to be considerable in order for nearby matter concentrations to adequately accelerate the MW over such relatively small scales.  On the other hand, if the LG's peculiar velocity is induced by more distant structures, then we know that anisotropies in the large-scale matter distribution must exist to at least those structures and that the universe becomes isotropic only at larger distances.  

To answer these questions, much effort has been spent examining the peculiar velocity that would be induced onto the LG by the distribution of various mass tracers, such as galaxies and clusters of galaxies, and comparing this motion with that inferred from the CMB.  Traditionally such analyses have made use of the linear theory of gravitational instability, which dictates that the peculiar velocity of a reference frame can be related to the gravitational acceleration induced by the mass distribution surrounding it via (Peebles 1976, see our Appendix A) 

\begin{equation}
\textit{\textbf{v}}_{p} \hspace{.1in} = \hspace{.1in} \frac{H_{o}\beta}{4\pi \bar{n}} \int \frac{n(r)}{r^{2}} \textit{\textbf{\^{r}}}\ dr 
\end{equation}

\noindent where $\beta = \Omega^{0.6}_{0}/b$ and $b$ is the biasing factor relating the mass-tracers to the underlying mass distribution they represent, and $\bar{n}$ is the average mass-tracer number density.  In other words, Equation 1 tells us that the dipole moment of a mass-tracer distribution can be directly related to the peculiar velocity that sample would induce on the LG.  Within this framework, the comparison of the LG's peculiar velocity as inferred from the CMB dipole anisotropy to that produced by a mass-tracer distribution can shed light on the role of the sample on producing the LG's motion, provide an estimate of the depth at which inhomogeneities in the distribution of the sample affect the LG's dynamics (i.e. the convergence depth, $R_{conv}$), and place constraints on how the sample traces the underlying matter distribution in the form of the biasing parameter $\beta$.  Performing this type of analysis is a nontrivial matter for numerous reasons: (i) for linear perturbation theory to be applicable, the dipole moment of the sample under study must be relatively well aligned with the observed peculiar velocity of the LG; (ii) the characteristic depth of the sample must be larger than the depth at which anisotropies in the sample can influence the LG's dynamics ($R_{*} > R_{conv}$); (iii) the number of mass-tracers must be sufficient to accurately sample the underlying density field and avoid the introduction of shot-noise errors. 

Wary of these concerns, this type of dipole analysis has been extensively applied to the LG.  While studies have used samples ranging from optical galaxies (Lahav, Rowan-Robinson \& Lynden-Bell 1988;  Lynden-Bell, Lahav \& Burstein 1989; Shaya et al. 1992; Hudson 1993) to X-ray selected AGN (Miyaji \& Boldt 1990), much of the early dipole work focused on IRAS galaxies due to their considerable sky coverage (Meiksin \& Davis 1986; Yahil, Walker \& Rowan-Robinson 1986; Strauss \& Davis 1988; Yahil 1988; Rowan-Robinson et al. 1990, 1991; Strauss et al. 1992; Plionis, Coles \& Catelan 1993; Basilakos \& Plionis 1998).  Despite the wide employment of such galaxy samples, they are all plagued by steeply declining selection functions with distance, leading to a significant incompleteness at large distances and characteristic depths of 80-100 $h^{-1}$ Mpc.  Analyses of galaxy samples have produced dipoles that are generally well aligned with the CMB dipole direction, but many differ regarding their implied convergence depth.  Strauss et al. (1992) and Hudson (1993) found an $R_{conv}$ of roughly $\sim$$50 h^{-1}$ Mpc, implying that all of the peculiar acceleration of the LG is induced relatively nearby.  On the other hand, Plionis (1988), Plionis, Coles \& Catelan (1993), Basilakos \& Plionis (1998), Branchini et al. (1999), Schmoldt et al. (1999), and Plionis et al. (2000) found evidence for contributions ranging from 100 to 150 $h^{-1}$ Mpc. The credibility of many of these early results must be questioned since the convergence depth of each sample shows a strong dependence on the sample's characteristic depth, a feature to be expected if anisotropies beyond that characteristic depth continued to contribute to the motion of the LG.  This implies that galaxy samples alone do not probe the mass fluctuation field deep enough to account for all the anisotropies that affect the LG's dynamics.     

The use of clusters of galaxies overcomes some of the problems faced by galaxy samples since clusters are luminous enough for samples to be volume-limited out to larger distances.  Although clusters only sparsely sample the underlying density field, which introduces shot-noise errors, they trace the peaks of the density fluctuation field, which have the greatest effect on the dipole amplitude and direction.  Using the Abell/ACO cluster catalogue (Abell 1958, Abell, Corwin \& Olowin 1989, hereafter ACO), which has a characteristic depth of over 240 $h^{-1}$ Mpc, evidence has been found for contributions to the LG dipole from depths of $\sim$160 $h^{-1}$ Mpc (Scaramella, Vettolani \& Zamorani 1991,1994; Plionis \& Valdarnini 1991; Brunozzi et al. 1995; Branchini \& Plionis 1996).  These results were confirmed by Plionis \& Kolokotronis (1998) using the X-ray Brightest Abell-type Cluster catalogue (XBACs, Ebeling et al. 1996), which is optically selected, but X-ray confirmed, thus eliminating projection effects which may have enhanced the dipole amplitude obtained with Abell/ACO (Sutherland 1988).

In the context of the dipole analysis, the primary limitation of the XBACs and Abell/ACO catalogues are their incompleteness at low Galactic latitudes.  This is because traditional optical cluster searches suffer from severe extinction and stellar obscuration in the direction of the Milky Way (MW), leading to catalogues with poor coverage in a $40^{\circ}$ wide strip centered on the plane of the Galaxy, known as the Zone of Avoidance (ZOA).  This is particularly troubling since Shaya suggested as early as 1984 that large-scale structures obscured by the ZOA could have a significant effect on the peculiar motion of the LG.  More recently, several studies have found renewed evidence for a significant bulk motion toward a vertex behind the plane of the MW (Riess et al 1997; Hudson et al. 1999), rekindling the idea of a Great Attractor (Lynden-Bell et al. 1988).  A variety of techniques have been used to reconstruct the ZOA, ranging from a uniform filling (Strauss \& Davis 1988; Lahav 1987) to a spherical-harmonics approach which extends structures above and below the plane into the ZOA (Plionis \& Valdarnini 1991, cf. Brunozzi et al. 1995).  The value of these reconstruction techniques is, however, limited if the MW does indeed obscure dynamically significant regions, as has been suggested.

With the recent compilation of the X-ray selected CIZA cluster catalogue (named for Clusters in the Zone of Avoidance; Ebeling, Mullis \& Tully 2002), it has become possible to add real cluster data to the region behind the MW.  In this study we combine the CIZA and XBACs samples, with appropriate weightings, to produce the first all-sky catalogue of X-ray luminous clusters and provide a dipole analysis of the LG peculiar velocity without the use of reconstruction methods to fill the ZOA.  We proceed in the following manner: in section 2 we give an overview of the XBACs and CIZA samples, with a thorough look at the systematic effects in each. Section 3 describes the details of the dipole analysis and our results are put forward in section 4.  Finally we summarize our primary conclusions in section 5.  Throughout this paper we assume an Einstein-de Sitter universe with $q_{0}=0.5$ and $H_{0}=100$ $h$ km s$^{-1}$Mpc$^{-1}$, so that our results are directly comparable to those of previous studies.

\begin{figure}[t]
\hspace*{-0.15in}
\centerline{\psfig{file=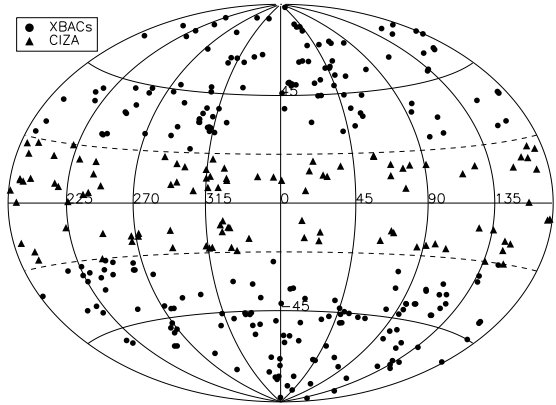,width=3.25in,angle=0}}
\caption{Aitoff projection of the XBACs and CIZA cluster catalogues in Galactic Coordinates.  The dashed lines represent the traditional Zone of Avoidance ($|b|<20^{\circ}$).}
\end{figure}

\section{Samples \& Systematics}
In this study we use the XBACs and CIZA X-ray cluster catalogues to estimate the gravitational acceleration induced by clusters on the LG.  The X-ray properties of these catalogues make them preferable over optical cluster catalogues such as Abell/ACO which have a high risk of corruption from projection effects (Lucey 1983; Sutherland 1988; Struble \& Rood 1991) and suffer from incompleteness at low Galactic latitudes.  The advantages presented by X-ray observations overcome both of these problems:  (i) cluster X-ray emission originates from the $\sim10^{7}$ K intracluster medium, which is more peaked at the gravitational center of the cluster than the projected galaxy distribution, thus requiring that clusters be in almost perfect alignment to be mistaken for a single, more luminous object; (ii) X-ray emission does not suffer as severe an extinction in the plane of the Galaxy, therefore allowing for the compilation of cluster catalogues in regions that optical searches have largely avoided.  In this section we describe the attributes of both the XBACs and CIZA cluster catalogues in further detail.

\begin{figure}[t]
\centerline{\psfig{file=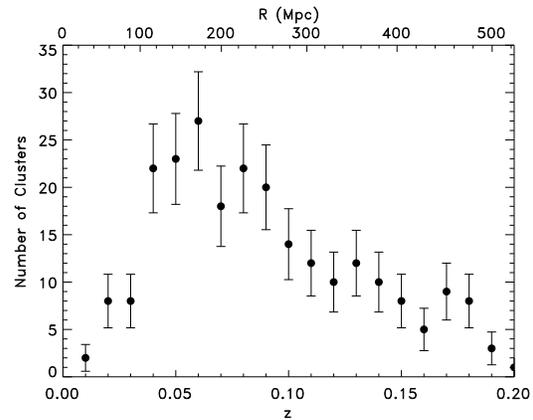,width=3.0in,angle=270}}
\caption{The redshift distribution of the 242 XBACs clusters with $f_{x} \geq 5\times 10^{-12}$ erg cm$^{-2}$ s$^{-1}$ in the 0.1--2.4 keV band, $z<0.2$ and $|b|>20^{\circ}$. The redshift binwidth is 0.01}
\end{figure}

\subsection{The XBACs Catalogue}
The XBACs sample has its origins in the ROSAT All-Sky Survey (RASS, Voges 1992) and consists of 242 clusters with X-ray fluxes greater than $5\times 10^{-12}$ erg cm$^{-2}$ s$^{-1}$ in the 0.1--2.4 keV band.   Although XBACs is X-ray confirmed, it is not X-ray selected, in that targets investigated in the RASS were selected from the Abell/ACO catalogue, thus the statistical XBACs sample is limited to Galactic latitudes of $|b|>20^{\circ}$ and redshifts of $z\leq 0.2$.  As discussed in Ebeling et al. (1996) (hereafter E96), the RASS data were initially processed using the Standard Analysis Software System (SASS, Voges et al. 1992) and cross-correlated with the Abell/ACO cluster catalogue.  Fields of interest were reprocessed with the Voronoi Tesselation and Percolation (VTP, Ebeling 1993; Ebeling \& Wiedenmann 1993) algorithm because of the difficulty of measuring extended emission with SASS.  Of the 10,241 SASS sources investigated, 278 were confirmed as clusters; 12 of these were at low Galactic latitudes ($|b|<20^{\circ}$) and 24 had redshifts greater than 0.2, leaving the 242 clusters we use in our analysis.  Unlike the Abell/ACO sample, this sample is not affected by a volume incompleteness as a function of distance for richness class $R=0$ clusters (E96).  Estimated to be 80\% complete, XBACs is the largest near-all-sky X-ray cluster catalogue published to date. The distribution of the sample in Galactic coordinates and redshift space is shown in Figures 1 and 2, respectively.

\subsection{The CIZA Catalogue}
Unlike XBACs, the CIZA survey has compiled a truly X-ray selected sample of clusters distributed behind the plane of the Galaxy.  As described in Ebeling, Mullis \& Tully (2002, hereafter EMT), CIZA targets were selected from the RASS Bright Source Catalogue ( Voges et al. 1999) if they met three criteria: (i) location in the ZOA, $| b | < 20^{\circ}$, (ii) an X-ray flux greater than $1\times 10^{-12}$ erg cm$^{-2}$ s$^{-1}$ (0.1--2.4 keV) and (iii) a spectral hardness ratio exceeding a preset threshold value\footnote{The minimum hardness ratio threshold depends on location in the plane; see EMT for details.} to discriminate against softer, non-cluster sources.  The use of the Bright Source Catalogue as a target list means CIZA is not correlated with any optically selected catalogues and contains clusters discovered by their X-ray properties alone.

\begin{figure}[t]
\centerline{\psfig{file=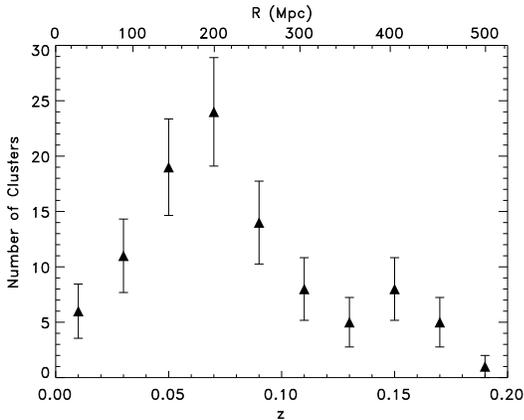,width=3.0in,angle=270}}
\caption{The redshift distribution of the 104 CIZA cluster with $f_{x} \geq 5\times 10^{-12}$ erg cm$^{-2}$ s$^{-1}$ in the 0.1--2.4 keV band, $z<0.2$ and $|b|<20^{\circ}$.  The redshift binwidth is 0.02}
\end{figure}

A subsample of 73 clusters with \textit{detect} fluxes above $5\times 10^{-12}$ erg cm$^{-2}$ s$^{-1}$ (the B1 sample) has recently been published (EMT).  Ebeling et al. (1998, hereafter E98) have shown these detect fluxes to be systematically low for nearby clusters due to the SASS algorithm's inability to reliable measure fluxes for extended objects.  Since the B1 sample is detect-flux limited, it is therefore not flux limited in the same sense as XBACs.  To allow the combination of CIZA and XBACs, we use the total fluxes of the B1 sample listed by EMT and also add clusters from the yet unpublished B2 and B3 CIZA subsample\footnote{The B2 and B3 subsamples are defined by detect fluxes greater than $3\times 10^{-12}$ and $2\times 10^{-12}$ erg cm$^{-2}$ s$^{-1}$, respectively.}, whose total fluxes within a metric 1 Mpc aperture exceed $5\times 10^{-12}$ erg cm$^{-2}$ s$^{-1}$, so as to compile a catalogue that is truly flux limited to the same level as XBACs.  The resulting sample, limited to $z<0.2$ and $|b| < 20^{\circ}$, contains 104 clusters, of which 73 come from the published B1 sample. The positions of these clusters in Galactic coordinates are shown in Figure 1 and their redshift distribution is shown in Figure 3.

\subsection{Systematics}
Although XBACs and CIZA are influenced by different systematic effects, we assume that both sample the same underlying cluster distribution.  In this section we discuss the creation of weights for each cluster in our sample which are meant to correct for any incompleteness present within our luminosity and redshift range.  Much of this analysis will be carried out by comparing the properties of XBACs and CIZA to those of the Brightest Cluster Sample (BCS, E98) which is the most complete, truly X-ray selected catalogue produced from the RASS to date, but due to its sky coverage (restricted to $\delta \geq 0^{\circ}$), is of limited value in the dipole analysis.


\begin{figure}[t]
\centerline{\psfig{file=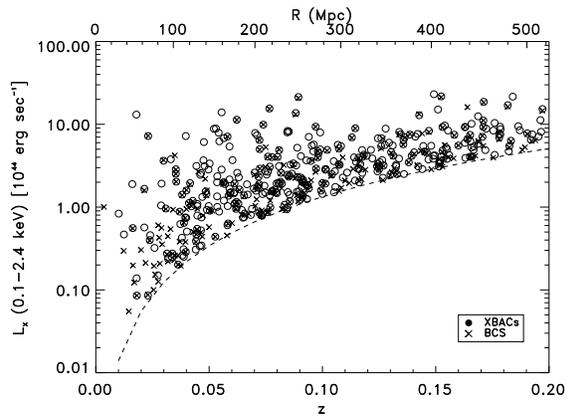,width=3.0in,angle=90}}
\caption{X-ray luminosity versus redshift for the XBACs and BCS cluster samples.  The dashed line represents the $3\times 10^{-12}$ erg cm$^{-2}$ s$^{-1}$ flux limit.  Note the disparity in the number of XBACs versus BCS clusters at low redshift/low luminosity}
\end{figure}

Many of the systematics in XBACs grow out of its nature as an optically selected sample; although X-ray confirmation effectively eliminates projection effects, clusters missed by Abell/ACO will not be included in XBACs.  Other biases are introduced by the RASS reduction algorithm's inability to accurately measure very extended emission; out of the 49 Abell clusters within a redshift of 0.05, only 30 are detected by SASS (E98).  To quantify these selection effects, we compare the number of XBACs and BCS clusters in luminosity-redshift space; this is shown in Figure 4.  We can resolve two forms of incompleteness in XBACS: (i) very nearby, very extended clusters are systematically missed; (ii) optically poor clusters of moderate X-ray luminosity are missed at all redshifts.  Evidence of the first effect is clearly apparent in the distribution of the BCS versus that of XBACs in Figure 4. 


To quantify and correct for both effects, we fit a fourth-order, two-dimensional polynomial to the ratio of the number of XBACs to BCS clusters in luminosity-redshift parameter space, from which we obtain corrective weights for each XBACs cluster.  These weights effectively represent the number of clusters that we estimate have been missed by XBACs due to its various selection biases.  We use these weights at all stages of the dipole analysis to compensate for the inherent incompleteness of the XBACs sample.  The corrected XBACs redshift distribution is shown on Figure 5 along with the BCS and unweighted XBACs samples.  As a consistency check, we note that the fitted luminosity functions, which are known to differ for the XBACs and the BCS samples (Ebeling et al. 1997, 1998), have consistent parameters when we employ our corrective weights, which one would expect if XBACs has been corrected in the proper manner.

\begin{figure}[t]
\centerline{\psfig{file=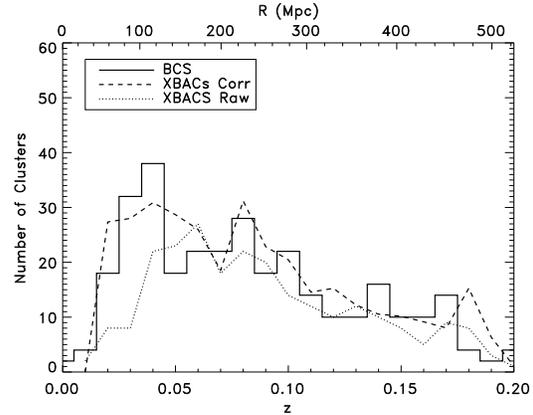,width=3.0in,angle=270}}
\caption{The redshift distribution for BCS clusters with $F_{x} \geq 5\times 10^{-12}$ erg cm$^{-2}$ s$^{-1}$ in the 0.1-2.4 keV band, $z<0.2$ and $b>20^{\circ}$.  Also plotted is the uncorrected and corrected XBACs samples.}
\end{figure}

We perform a similar analysis on CIZA, but since the sample is X-ray selected, it matches the BCS far better than XBACs does, with the catalogue's primary incompleteness coming from severe extinction within $\pm5^{\circ}$ of the Galactic plane. EMT argue that this incompleteness is correlated with distance and significant only at higher redshifts ($z>0.075$), due to the increasing difficulty of obtaining spectroscopic confirmation of distant clusters in the very central regions of the Galactic plane.  Therefore CIZA is statistically complete for nearby clusters which have an increased weight on the dipole amplitude.  Since we limit our dipole analysis to distances within 240 $h^{-1}$ Mpc ($z\sim 0.08$), a redshift range EMT argue is statistically complete, we apply no corrective weights to the individual CIZA clusters.

\section{The Dipole Methodology}
In the most general sense, the monopole and dipole moments of a distribution of objects are given by

\begin{equation}
M =\sum_{i=1}^{N} w_{i}; \hspace{0.25in} 
\textit{\textbf{D}} = \sum_{i=1}^{N} w_{i}\textit{\textbf{\^{r}}}_{i},
\end{equation}
\noindent
where $w_{i}$ are individual weights and $\textbf{\^{r}}_{i}$ are the unit vectors pointing to the position of each object.  If the dipole is to represent the peculiar gravitational acceleration acting on an object at the origin, $w_{i}$ needs to contain the inverse square component $1/r^{2}$. In the case of a flux-limited catalogue, objects require an additional weighting to correct for the non-detection of intrinsically less luminous members with increasing distance.  This is achieved by weighting each object with the inverse of the sample's selection function, $\phi(r)$ (see Equation 9), which represents the fraction of objects that are observed above the sample's flux limit at some distance $r$.  Other weights include the mass of the object, $M_{i}$, and corrections for any known systematic effects. Equations 2 can be rewritten as

\begin{equation}
M =\sum_{i=1}^{N} \frac{w_{i}}{\phi(r_{i})r_{i}^{2}}; \hspace{0.15in} 
\textit{\textbf{D}} = \sum_{i=1}^{N} \frac{w_{i}}{\phi(r_{i})r_{i}^{2}}\textit{\textbf{\^{r}}}_{i},
\end{equation}

From these equations it can be seen that the characteristics of the dipole are such that its amplitude will increase with distance until the largest inhomogeneity in the sample is encompassed and isotropy is reached, after which the dipole flattens out to its final value. This flattening signals that the convergence depth, $R_{conv}$, has been reached.  The monopole amplitude, on the other hand, increases linearly with distance for a uniform distribution and is related to the average tracer density as $M(R) = 4\pi\bar{n}R$, as long as $R > R_{conv}$. 

With this definition of the dipole, we can use linear perturbation theory with the assumption of linear biasing (Equation 1) to relate the peculiar velocity of an observer to a surrounding noncontinuous mass-tracer sample.  A more useful version of Equation 1 can now be written as

\begin{equation}
\textit{\textbf{v}}_{p} = \frac{H_{o}\beta}{4\pi \bar{n}} \hspace{.05in} \sum_{i=1}^{N} \frac{w_{i}}{\phi(r_{i})r_{i}^{2}}\textit{\textbf{\^{r}}}_{i}
\end{equation}
\begin{eqnarray}
\hspace{.55in} = \hspace{.1in} \beta \textit{\textbf{D}}_{tracer} \hspace{0.28in} \nonumber
\end{eqnarray}
where $w_{i}$ represent any additional weights (see \S2.3) and $\textit{\textbf{D}}_{tracer}$ is the measurable dipole.  Note we have folded the constants $H_{o}/4\pi \bar{n}$ into $\textit{\textbf{D}}_{tracer}$; it is this value that we refer to as the dipole throughout the rest of the paper.

In the case of the LG, $\textit{\textbf{v}}_{p}$ has been measured from the CMB dipole to be $627\pm 22$ km/s in the direction of $(l,b) = (276^{\circ}, 30^{\circ})$ (Kogut 1993) and in this study we measure $\textit{\textbf{D}}_{cl}$ from the cluster distribution.  Assuming that $\textit{\textbf{v}}_{p}$ is well aligned with the velocity predicted from the cluster dipole, this relation provides the means to estimate $\beta$ ($=\Omega^{0.6}_{0}/b$).  As we show later, the dipole of the XBACs+CIZA cluster sample is sufficiently aligned with the CMB dipole to warrant the use of linear theory and therefore Equation 4.  We use this relation throughout the remainder of this paper.

\subsection{Weights}
The primary weight used in the dipole analysis is that of distance.  All of the clusters in our sample have measured heliocentric redshifts, $z_{\odot}$, which we transform into the LG and CMB rest frames using

\begin{equation}
                            cz_{_{LG}} = cz_{\odot}+300\sin l \sin b
\end{equation}
and
\begin{eqnarray}
  cz_{_{CMB}} = cz_{_{LG}}+v_{_{LG}}\hspace{0.02in}\big[\hspace{0.02in}\sin(b)\sin(b_{_{CMB}})\\ +\hspace{0.05in} \cos(b)\cos(b_{_{CMB}})\cos(|l_{_{CMB}}-l|)\hspace{0.01in}\big] \nonumber
\end{eqnarray}
where $v_{_{LG}}$ is the amplitude of the LG velocity as inferred from the CMB dipole anisotropy and $(l_{_{CMB}},b_{_{CMB}})$ is the direction of this motion in Galactic coordinates (after correcting for virgocentric infall, see \S4).  All redshift are then converted into distances using Mattig's formula (1958):
\begin{equation}
        \nonumber    r = \frac{c}{H_{0}q_{0}^{2}(1+z)}[q_{0}z+(1-q_{0})(1-\sqrt{2q_{0}z+1})]
\end{equation}
which reduces to 
\begin{equation}
	             r = \frac{2c}{H_{0}}(1-\frac{1}{\sqrt{z+1}})
\end{equation}
for our assumed value of $q_{0} = 0.5$.  As is described in \S3.2, we calculate the cluster dipole in both the LG and CMB rest frames since it has been shown that they over- and under-estimate, respectively, the real-space dipole obtained using detailed reconstruction techniques (Branchini \& Plionis 1996).  To avoid shot-noise effects which arise out of the sparse sampling of clusters at higher redshifts and to facilitate comparisons with the literature, we limit our dipole analysis to a distance of 240 $h^{-1}$ Mpc.

The next component of our weighting scheme takes into account the contribution of each cluster to the dipole in terms of its relative mass.  We estimate the mass of each cluster in our sample through the empirical relationship $M \propto L^{3/4}_{x}$ (Allen et al. 2003) which links a cluster's X-ray luminosity to its mass contained within the radius $R_{200}$, defined as the distance where the mean enclosed density is 200 times the critical density of the universe at the redshift of the cluster.  Due to the steep shape of the X-ray luminosity function, this weight is essential so as to prevent low-luminosity systems from artificially dominating the dipole signal by their sheer number.  

The final weight given to all clusters is the inverse of the sample's selection function at the cluster distance, which is meant to correct for the non-detection of intrinsically less luminous clusters with increasing distance.  The selection function is defined as the fraction of the cluster number density that is observed above the flux limit at a given distance:
\begin{equation}
   \phi(r) = \frac{1}{{\bar{n}_{c}}} \int^{\infty}_{L_{min}(r)} \Phi_{X}(L)dL,
\end{equation}
where $\bar{n}_{c}$ is the average cluster density, $\Phi_{x}(L)$ is the X-ray cluster luminosity function and $L_{min}(r) = 4\pi r^{2}S_{lim}$, where $S_{lim}$ is the flux limit.  We estimate $\bar{n}_{c}$ by integrating the luminosity function over the entire luminosity range of the sample

\begin{equation}
                 \bar{n}_{c}  = \int^{\infty}_{L_{min}} \Phi_{X}(L)dL,
\end{equation}
where $L_{min}$ is set to $1.25\times 10^{42}$ ergs s$^{-1}$ (E97, E98).  The value calculated for $\bar{n}_{c}$ is used to normalize the XBACs+CIZA dipole as well as determine a value of the selection function. 

For the luminosity function, $\Phi_{X}(L)$, we adopt a single Schechter-like function of the form (E98)
\begin{equation}
            \Phi_{x}(L) = A \exp\left(-\frac{L}{L_{*}}\right)L^{-\alpha}   
\end{equation}
to describe the \emph{true} cluster X-ray luminosity distribution. We further assume this \emph{true} distribution to be best represented by the BCS, since it is the most complete, truly X-ray selected catalogue produced from the RASS and adopt the description presented in E97.  Since we have already corrected for any incompleteness in our samples (see \S2.3) we use this single luminosity function to describe our combined X-ray cluster data set.

\subsection{Cluster Peculiar Velocities}
The peculiar velocity of clusters may alter the perceived distance to a cluster and hence its contribution to the dipole signal.  Two primary methods have been employed in the literature to correct for this effect; the first is to reconstruct the bulk motion of clusters in the local universe and subtract the estimated peculiar velocity from each cluster redshift in your sample (Branchini \& Plionis 1996).  The second method involves computing the dipole in both the LG and CMB reference frames and assuming that these over- and under- estimate, respectively, the real-space dipole (Brunozzi et al. 1995).  The reasoning for this latter method is as follows: in the LG frame the peculiar velocities of nearby clusters is negligible since they participate in the local bulk flow with us (see Dekel 1994 for a review), but more distant systems that are detached from this motion will appear closer to us if they lie in the direction of our motion.  This effectively enhances the dipole signal these distant clusters would normally contribute and artificially increases the dipole amplitude.  On the other hand, placing us in the CMB frame removes the LG's motion from the redshifts of these distant cluster, but also detaches us from the local bulk flow, making nearby clusters appear farther than they really are.  The dipole signal of these clusters is reduced and hence the dipole amplitude is artificially decreased.   Since the LG frame overestimates the real-space dipole and the CMB frame underestimates it, the assumption is made that the true dipole lies between these two extremes.

Since the reconstruction method of Branchini \& Plionis (1996) requires an a priori estimate of the $\beta$ parameter, a value we wish to measure, we have decided to use the latter of these two methods to account for the peculiar velocity of clusters and calculate the cluster dipole in both the LG and CMB frames.

\begin{figure*}
\vspace*{-4mm} \mbox{}\\
\epsfxsize=6in 
\epsffile{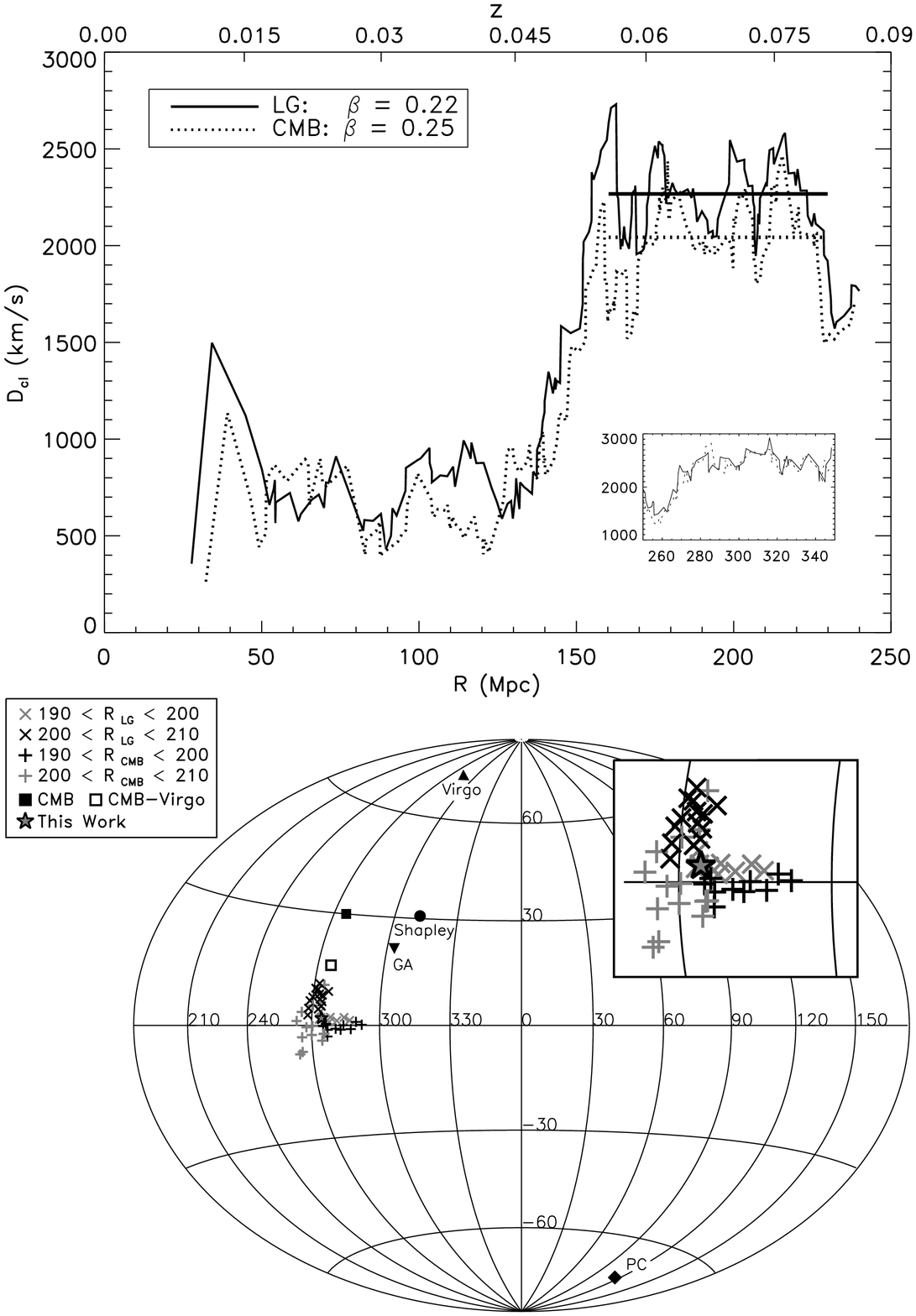}\\

\baselineskip0.6mm
Fig.~6.--- The X-ray cluster dipole amplitude versus distance ($above$) and the dipole direction in Galactic coordinates ($below$).  The amplitude is derived in both the LG and CMB frames, since calculations in each frame overestimate and underestimate, respectively, the true cluster dipole.  The filled square is the nominal CMB pointing of ($l$,$b$) = (276$^\circ$,30$^\circ$), while the open square is the CMB pointing corrected for a 170 km/s Virgocentric infall velocity.  The star represents the average cluster dipole direction, which is an average of the median direction in the LG and CMB frames between 190 and 210 $h^{-1}$ Mpc. \\
\end{figure*}

\section{Results and Discussion}

\subsection{Dipole Amplitude and Direction}

The primary results of our analysis, specifically the dipole amplitude, direction and inferred biasing parameter are shown in Figure 6.  We find that (i) the cluster distribution becomes isotropic with respect to the LG at a distance of 160 $h^{-1}$ Mpc, as is indicated by the flattening of the dipole amplitude, (ii) the asymptotic value of the amplitude is largely set by the coherent signal near $\sim150 h^{-1}$ Mpc and (iii) there is good agreement in the dipole direction as measured by both the X-ray cluster distribution and the CMB anisotropy throughout our study volume.  These results are in agreement with the findings of previous analyses that utilized XBACs without the CIZA supplement (Plionis \& Kolokotronis 1998), suggesting that to some extent the newly discovered CIZA clusters trace the same large-scale anisotropy in the matter density distribution as do previously analyzed X-ray cluster catalogues.  On the other hand, our dipole profile differs from results obtained using a variety of galaxy catalogues that suggested a local convergence of the LG peculiar acceleration (Lynden-bell et al. 1989, Rowan-Robinson et al. 1990, Strauss et al. 1992, Hudson 1993), reaffirming that galaxy samples alone do not probe the mass fluctuation field deep enough to account for all the anisotropies that affect the LG's dynamics.  This fact was predicted by Peacock (1992) and Shaya et al. (1992) on theoretical and observational grounds, respectively, and contributions to the dipole from distances larger than the limiting volumes of many galaxy samples were measured in optical cluster catalogues by Scaramella et al. (1991, 1994), Plionis et al. (1993), but at lower levels than that observed in the X-ray cluster dipole.  Scaramella et al. (1991) estimate that less than $30\%$ of the LG's peculiar velocity comes from distances greater than $\sim 60 h^{-1}$ Mpc, while we find that roughly $\sim70\%$ of our dipole amplitude comes from contributions at similar distances.  This difference is partly due to the X-ray selected nature of our sample: clusters in the local neighborhood fail to be detected due to their highly extended X-ray emission and hence the dipole amplitude is depressed and the resulting relative contribution from more distant clusters is enhanced.  We further find that most of the signal originating at distances greater than $100 h^{-1}$ Mpc comes from the most dramatic feature in the dipole profile: the steep rise at roughly 150 $h^{-1}$ Mpc.  As is discussed in \S4.1, this increase is largely due to the Shapley concentration (Shapley 1930) of clusters which lie at roughly this distance.  This feature displays quite dramatically that a significant amount of the peculiar acceleration acting on the LG is induced from distances greater than those sampled by many galaxy catalogues.

\begin{figure*}[t]
\vspace*{-4mm} \mbox{}\\
\hspace*{3in} \\
\epsfxsize=6in 
\epsffile{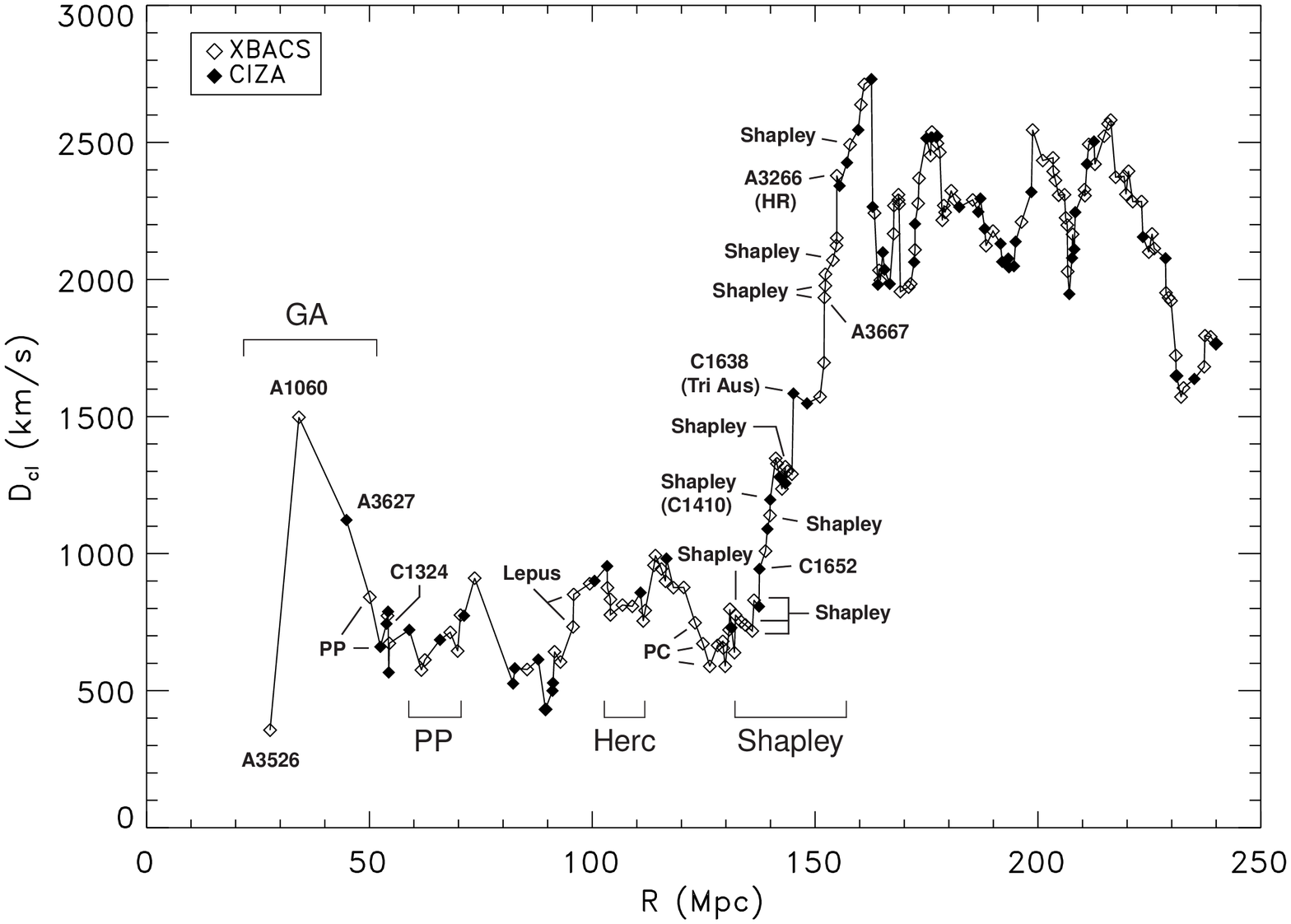}\\
\baselineskip0.6mm
Fig.~7.--- Schematic dipole profile; see text for details. Each symbol represent a cluster used in our analysis.  Abell and CIZA clusters begin with the letters 'A' and 'C', respectively.  Acronyms are GA: Great Attractor, PP: Perseus-Pegasus, PC: Pisces-Cetus, HR: Horologium-Reticulum.  We find that the Shapely concentration is the single supercluster most responsible for producing the increase in the dipole signal between 140 and $160 h^{-1}$ Mpc. 
\end{figure*}

Also noteworthy is the good agreement between the cluster and CMB dipole directions: using the XBACs+CIZA sample, we find the average cluster dipole in the LG and CMB frames at 200 $h^{-1}$ Mpc to be misaligned from the CMB dipole direction by $14^{\circ}$.  With the addition of the Virgo cluster, this misalignment angle can be reduced to within $\sim3^{\circ}$ (we discuss Virgo in greater depth in \S4.3).  Also interesting is that the dipole signal produced by the nearest XBACs+CIZA clusters points only $\sim10^{\circ}$ away from the overall CMB dipole pointing.  This indicates, as did previous studies (Scaramella et al. 1991, Basilakos \& Plionis 1998), that our dipole direction is roughly aligned with the CMB at fairly shallow distances, suggesting a coherence in the X-ray cluster dipole thought to be brought about by the alignment of large-scale structures toward the CMB anisotropy.  The nature of these structures will be examined in the next section, but it is apparent that early studies which found evidence for a local convergence of the dipole were misled due to this early alignment and the fact that the dipole profile remains temporarily flat out to distances well beyond the limiting depth of many galaxy catalogues.





\subsection{The Dipole Profile}

Our dipole profile differs from those obtained using optical cluster samples such as Abell/ACO, which often produce profiles that rise to their asymptotic amplitudes fairly quickly and with little resolved structure (see Figure 2 of Brunozzi et al. 1995).  This difference is presumably due to the sparseness of our X-ray cluster sample; by selecting the most massive and therefore most dynamically significant clusters, we are able to resolve the contributions of specific groups of clusters, or superclusters, to the overall dipole amplitude.  In this section we examine the dipole profile in greater detail and link observed structures to the superclusters that produce them.

\begin{figure*}[t]
\vspace*{-4mm}
\hspace*{1in}
\mbox{}
\epsfxsize=4in 
\epsffile{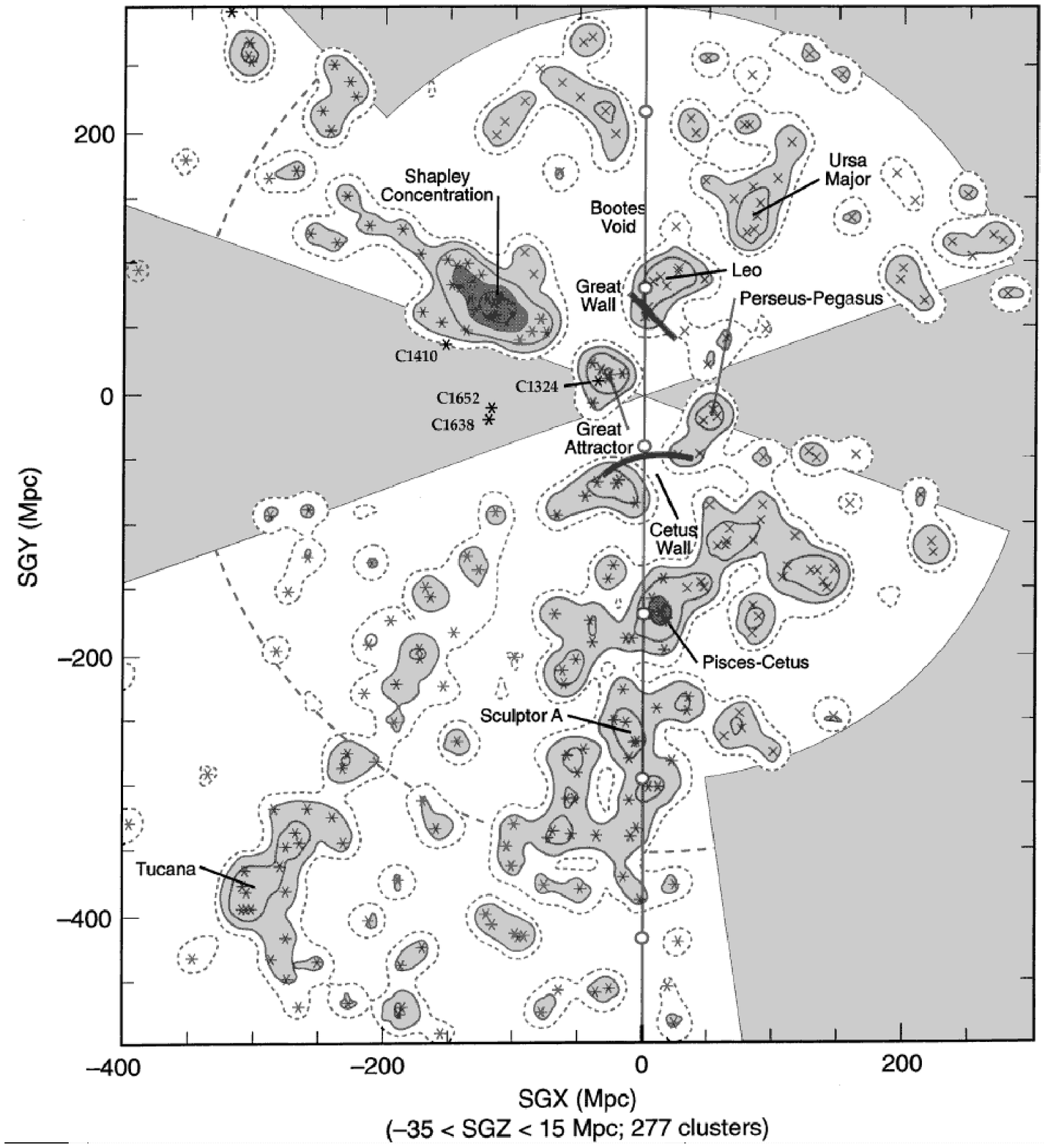}\\
\baselineskip0.6mm
\vspace{0.1in}\\
Fig.~8.--- An SGY-SGX projection of a region $50 h^{-1}$ Mpc thick in SGX.  The first and second contour shades steps are 1 and $3\times 10^{-3}$ clusters Mpc$^{-2}$.  The dashed contour is at a level of $0.5\times 10^{-3}$.  The traditional ZOA is the shaded region centered on SGY=0.  Notable structures such as the GA, PP, and Shapley regions are labeled, as well as the individual CIZA clusters C1324, C1652, C1638 and C1410.  Figure adapted from Tully et al. (1992); used with permission.
\end{figure*}

In Figure 7 we present a schematic version of the dipole profile in which we highlight dynamically significant clusters and associations.  The nearest cluster in our sample is A3526, followed by A1060 and A3627, all three of which lie in the traditional GA region (otherwise known as Hydra-Centaurus), which was originally proposed by Shaya (1984) and Lynden-Bell et al. (1988) and later by Dressler \& Faber (1990), Dekel (1994) and da Costa et al. (1996).  These three clusters are responsible for the first feature seen in our profile, the bump observed at roughly 40 $h^{-1}$ Mpc.  Following a drop largely due to two clusters opposite the GA, the amplitude increases as a result of the newly discovered massive cluster CIZA1324.6-5736 (hereafter we will refer to CIZA clusters only by their right ascension, i.e. C1324) which lies between A3526 and A3627 on the sky, but at a greater distance ($\sim 54 h^{-1}$ Mpc).  EMT suggest that the existence of this cluster indicates the GA may be more extended than previously thought and is perhaps best described as a wall-like, triangular structure with C1324, A3526 and A3627 at its vertices.  Also interesting is that the dipole signal produced by this trio is already in rough agreement with the CMB dipole pointing (within $\sim10^{\circ}$).  As mentioned in the previous section, this fact led many early studies using shallow galaxy samples to claim they had reached the convergence depth just beyond the GA region and conclude that the LG's peculiar motion was entirely induced by nearby mass concentrations (Strauss \& Davis 1988, Lynden-Bell et al. 1989, Strauss et al. 1992, Hudson 1993).  This conclusion was troubling since observations failed to detect the amounts of matter needed to produce such a local acceleration, a fact Shaya et al. (1992) pointed out and used as evidence for more distant contributions to the dipole.  From Figure 7 we can clearly see that significant amounts of the dipole signal originate at larger distances, but as we shall see shortly, the alignment of the GA region and the CMB pointing is not coincidental.
  
Following the GA region in Figure 7 comes a significant dip in the dipole signal which is primarily do to the inclusion of the Perseus-Pegasus (PP) supercluster (sometimes called Perseus-Pisces or just Perseus), which is directly opposite the GA, but slightly further away from the LG.  The polarity of these two structures is evident in Figure 8, which was adapted from Tully et al. (1992).  Although the ZOA is not filled in, Figure 8 (their Figure 4) highlights some of the most significant structures in the local neighborhood, including the GA, PP, Leo and Pisces-Cetus superclusters as well as the Shapley concentration.

Proceeding to greater distances, we encompass the three distinct cluster associations Lepus, Leo and Hercules, whose dynamical effects are manifested as the amplitude oscillations between 80 and 120 $h^{-1}$ Mpc.  Following the Hercules region of influence comes a decrease in the amplitude due to the nearest clusters in the highly extended Pisces-Cetus concentration.  This downturn is reversed shortly thereafter as we reach the first four clusters in the Shapley supercluster.

Originally recognized by Shapley (1930) and more recently by Scaramella et al. (1989) and Raychaudhury (1989), the Shapley supercluster is the most concentrated distribution of clusters in our X-ray selected sample and is the greatest single contributor to the dramatic increase in the dipole amplitude between 140 and 160 $h^{-1}$ Mpc.  Using a friends-of-friends algorithm provided by R.B. Tully (private communication) we have identified 13 clusters in our sample that can be associated with the traditional Shapely region, eleven of which lie between 140 and 160 $h^{-1}$ Mpc; this subset is highlighted in Figure 7.  The Shapley Concentration's possible significance to the peculiar motion of the LG was recognized by Scaramella et al. (1989,1991) and Vettolani et al. (1990) for the high concentration of Abell clusters in the region and because of its directional alignment with the GA, which can be seen in Figure 8.  This is the reason why the alignment of the CMB dipole and the GA region is not a coincidence nor the result of a solely local acceleration induced by the GA, instead it is the bootstrap effect of the aligned mass concentrations of the GA and Shapley regions that go a long way in setting the amplitude and direction of the peculiar motion of the LG.

In addition to the Shapley Concentration, we can pinpoint four massive clusters that contribute to the increase in the dipole amplitude between 140 and 160 $h^{-1}$ Mpc; these are the CIZA clusters C1652 and C1638\footnote{C1638 is a redetection of the well known Triangulum Australis cluster} and the ACO clusters A3667 and A3266.  The locations of the two CIZA clusters have been marked on Figure 8.  EMT speculate that C1652 and C1638, which lie fairly near to each other and well in the ZOA could have a significant effect on the dynamics of the LG.  Considering C1638 alone produces nearly a 300 km/s increase in the dipole amplitude, their supposition seems to be correct. 

Moving beyond 160 $h^{-1}$ Mpc we find that the amplitude oscillates about its median value of 2267 $\pm192$ km/s (LG frame) due to the incorporation of superclusters such as the central portions of Pisces-Cetus at $\sim$170 $h^{-1}$ Mpc and Horologium-Reticulum at $\sim$180 $h^{-1}$ Mpc.  The apparent dropping off of the dipole signal at 230 $h^{-1}$ Mpc is only temporary, with the amplitude increasing shortly thereafter and remaining flat out to at least 350 $h^{-1}$ Mpc (see inset of Figure 6).

\setcounter{figure}{8}

\begin{figure}[t]
\centerline{\psfig{file=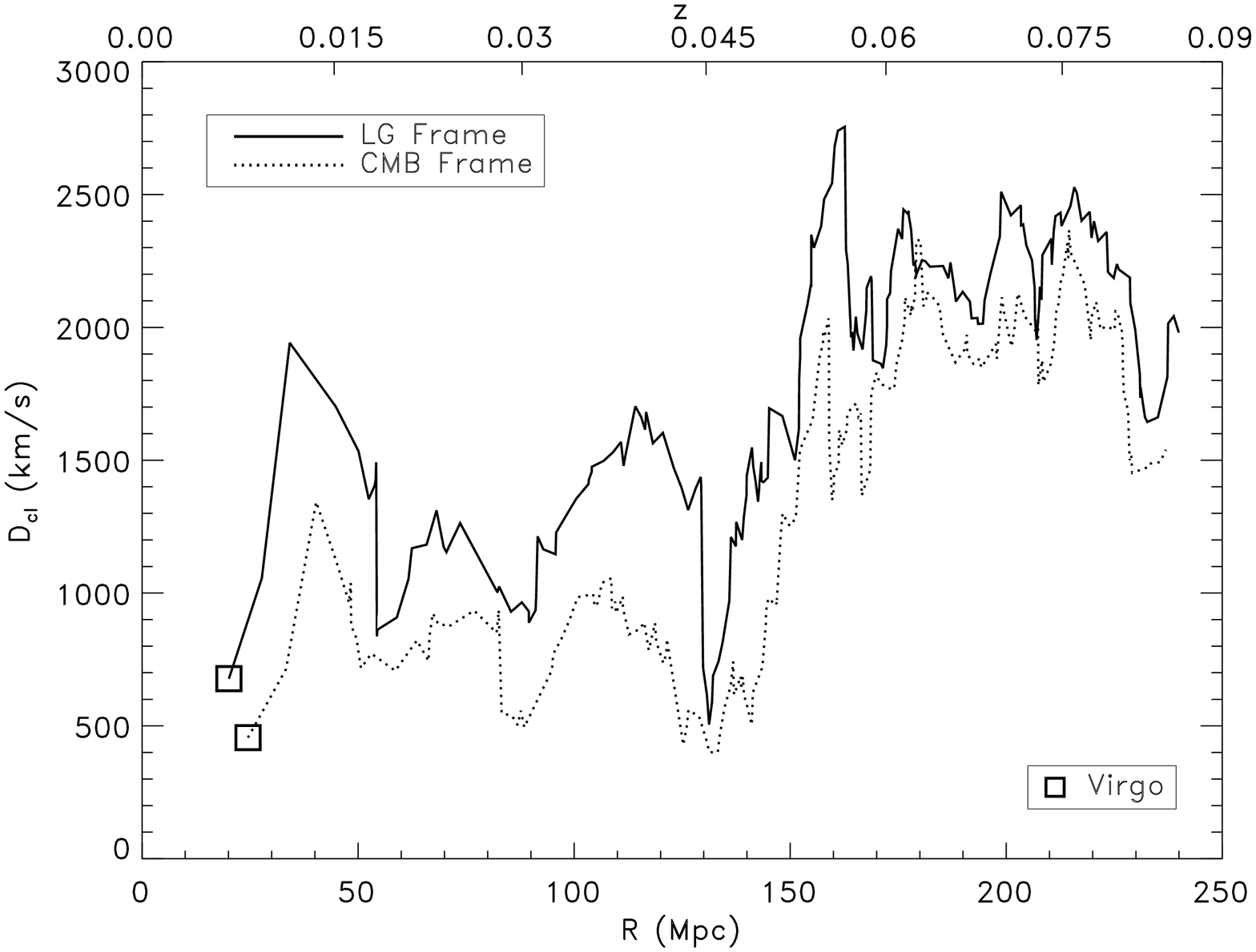,width=3.0in,angle=90}}
\caption{The XBACs+CIZA+Virgo cluster dipole amplitude versus distance in both the LG and CMB reference frames.  Virgo is weighted by a mass of $\sim3.5\times10^{14} M_{\odot}$ which is derived from its X-ray luminosity. See text for details.}
\end{figure}

\subsection{The Local Supercluster}

We have thus far excluded the Virgo cluster from our analysis since its large extent on the sky precluded its addition to the Abell catalogue and hence the XBACs catalogue as well.  Despite its absence from our sample, Virgo is known to produce a significant potion of the LG's peculiar velocity, therefore in this section we add the cluster to our sample and investigate its contribution to the final dipole amplitude and direction. 

Estimates of the dynamical effect of Virgo on the LG, through the measurement of the Virgocentric infall velocity, $v_{inf}$, has ranged from the minimal $v_{inf} = 85$ km/s (Faber \& Burnstein 1988), to the more significant $v_{inf} = 240$ km/s (Jerjen \& Tammann 1993).  Likewise the mass estimates of Virgo cover a large range: $\sim1\times10^{14} M_{\odot}$ within the central $1^{\circ}.5$ around M87 (Fabricant \& Gorenstein 1983),  $\sim3.5\times10^{14} M_{\odot}$ from the cluster's X-ray luminosity within a 1.8 Mpc radius (B\"{o}hringer 1994), $7\times10^{14} M_{\odot}$ from the infall pattern of galaxies near the cluster (virial theorem; Tully \& Shaya 1984) and $\sim1\times10^{15} M_{\odot}$ from both Least Action modeling (Shaya et al. 1995) and the requirement to produce a 250 km/s infall velocity at the LG distance (Davis et al 1980).  When we calculate the Virgo mass, and hence its weight in the dipole, using the empirical relationship $M \propto L^{3/4}_{x}$ of Allen et al. 2003, so as to be consistent with the other clusters in our sample, we obtain a mass of $1.8\times10^{14} M_{\odot}$.  Weighting Virgo by this mass and placing it at a distance of 21 Mpc (Tammann 1991) we find that the cluster has only a marginal impact on the dipole amplitude and direction: the final amplitude rises by $\sim100$ km/s, the relative contribution from large distances ($>60 h^{-1}$ Mpc) drops from $70\%$ to $60\%$ and the dipole pointing shifts to a higher Galactic longitude by $4^{\circ}$.  If we instead use B\"{o}hringer's (1994) larger estimate of $\sim3.5\times10^{14} M_{\odot}$ we find the dipole zero-point increases by 300 km/s, dropping the relative contribution from larger distances to $50\%$ (see Figure 9) and shifting the dipole pointing by $11^{\circ}$ .  If we center all of the mass which induces the LG's Virgocentric infall velocity, $\sim1\times10^{15} M_{\odot}$, to the position of Virgo, we find that contributions from clusters at distances greater than $60 h^{-1}$ Mpc reduces to $42\%$.  Further we find that the cluster dipole direction can be brought to within $3^{\circ}$ of the CMB dipole pointing when Virgo is weighted by a mass on the order of $10^{15} M_{\odot}$.

Adding Virgo to our analysis highlights two facts: first, even very modest incompleteness at very low redshift, such as the omission of Virgo from the XBACs catalogue, can have a dramatic effect on the dipole zero-point and hence the relative contributions from clusters at larger distances.  As we add mass to Virgo, we see the contribution from distances greater than $60h^{-1}$ Mpc consistently decrease.  Secondly, despite being misaligned with the CMB dipole by $45^{\circ}$, given an adequate mass Virgo plays an important role in setting the final cluster dipole direction.  When we examine the change in the dipole pointing with distance, we find four mass concentrations that dramatically alter the dipole direction:  the previously mentioned GA and Shapley regions, as well as the Virgo Cluster and the Pisces-Cetus (PC) supercluster; these four are highlighted in the lower panel of Figure 6.  Despite the obvious differences between Virgo and PC (the former a single, nearby cluster at a distance of $21h^{-1}$ Mpc, while the latter a much more massive association of clusters $130h^{-1}$ Mpc away) they have similar, but opposite effects on the dipole direction.  Whereas the Virgo cluster's position near the north Galactic pole pulls the dipole pointing toward higher Galactic longitudes, the PC concentration's more massive, yet more distant contribution near the south Galactic pole largely counteracts Virgo's effect and shifts the dipole toward lower Galactic longitudes.
Considering that the GA and Shapely regions lie about $35^{\circ}$ away from the CMB pointing, we conclude that it is the competing effects of the Virgo cluster and the Pisces-Cetus concentration that allow the GA and the Shapley supercluster to largely set, although not completely determine, the direction of the LG's peculiar motion.

\subsection{The $\beta$ Parameter}

With the direction of the X-ray-cluster and CMB dipoles fairly well aligned, we can use Equation 4 to find the biasing parameter, $\beta = \Omega^{0.6}_{0}/b$, by comparing the peculiar velocity of the LG as inferred from the CMB dipole, $\textit{v}_{p}$, to that predicted by the X-ray cluster distribution, $\textit{D}_{cl}$.  The caveat in this process again lies with the Virgo cluster which is not represented in our sample.  Traditionally authors have removed the infall velocity of the LG toward Virgo in order to have a meaningful comparison between the cluster and CMB dipoles.  We proceed in this manner, but also show results obtained by adding Virgo to our sample at the end of this section.

To obtain $\textit{D}_{cl}$, we calculate the median of the dipole amplitude between 160 and 230 $h^{-1}$ Mpc, where we assume isotropy has been reached and the amplitude has arrived at its final value.  Without Virgo, the LG and CMB reference frame amplitudes are $2267 \pm 192$ and $2043 \pm 229$, respectively, where the errors listed are from the variation of the amplitude over the given range.  

To remove the LG's Virgocentric infall velocity from its peculiar motion we use
\begin{equation}
	 \textit{v}^{\prime}_{p} = v_{p} - v_{inf}cos(\delta\theta)
\end{equation}
where $v_{p}$ is 627 km/s, $\delta\theta$ is the angle between the CMB dipole and Virgo directions and is roughly $45^{\circ}$ and we set $v_{inf}$, the infall velocity, to the literature average value of 170 km/s, which is employed by Plionis and Kolokotronis (1998) and used throughout much of the literature .  With these values we obtain a corrected peculiar velocity of $v^{\prime}_{p} = 507$ km/s.  Dividing this by the dipole amplitudes we obtain the following upper and lower estimates on the $\beta$ parameter:
\begin{eqnarray}
		\beta_{LG} = 0.22\pm 0.02 \nonumber \\
  		\beta_{CMB} = 0.25\pm 0.03\nonumber
\end{eqnarray}
where the errors are again from the variation of the amplitude over the distance of 160 and 230 $h^{-1}$ Mpc.




These estimates are in line with the $\beta = 0.24$ result of Plionis \& Kolokotronis (1998) who analyzed the XBACs sample using reconstruction techniques to fill the ZOA, but are slightly larger than most estimates using optical cluster samples.  Brunozzi et al. (1995) and Branchini \& Plionis (1996) obtain $\beta_{LG} = 0.15$ and $\beta = 0.21$, respectively, from their analysis of the Abell/ACO sample, while Plionis \& Valdarnini (1991) find $\beta_{LG} = 0.19$ and Scaramella (1991) finds $\beta_{LG} = 0.18$.  This could imply that X-ray bright clusters have an intrinsically greater mass-to-light bias or that the linear biasing assumption is not valid in the densest regions of the local universe, such as the Shapley supercluster, which dominate the X-ray cluster dipole. 

It should be noted that when we use the more traditional value of 240 km/s for the Virgocentric infall velocity, our average $\beta$ estimate drops to $0.21\pm 0.02$.  In addition, when we include Virgo we obtain $\beta$ values of $0.30\pm0.03 $ and $0.28\pm0.03$ for our lower, $1.8\times10^{14} M_{\odot}$, and higher, $1\times10^{15} M_{\odot}$, mass weightings, respectively.

\section{Conclusions}

One of the limitations to the use of clusters of galaxies to identify the structures inducing the LG's peculiar motion has been the incompleteness of cluster catalogues at low Galactic latitudes.  In this study we have combined the recently published CIZA catalogue with the XBACs cluster sample to produce the first all-sky catalogue of X-ray bright clusters in order to analyze the origin of the LG peculiar velocity without the need for reconstruction methods to fill the traditional ZOA.  We find that the cluster distribution becomes isotropic with respect to the LG at a distance of 160 $h^{-1}$ Mpc and that with Virgo excluded the asymptotic value of the dipole amplitude is largely set by a coherent signal near $\sim150 h^{-1}$ Mpc.  While this agrees with previous results, our finding that $\sim70\%$ of the dipole amplitude is set at distances larger than $60 h^{-1}$ Mpc differs from results obtained using galaxy and optically selected cluster samples.  By examining the dipole profile on a cluster-by-cluster basis, we conclude that the Shapley concentration is the single supercluster most responsible for producing the dipole signal between 140 and 160 $h^{-1}$ Mpc.  We also find that the cluster dipole is fairly well aligned with the direction of the CMB dipole at relatively shallow distances, and that the competing effects of Virgo cluster and the PC concentration essentially counteract each other's effects on the final dipole pointing.  These facts, coupled with the significant contributions to the LG peculiar acceleration from larger distances, reaffirm the bootstrap theory which suggests that the aligned mass concentrations of the GA and Shapley regions largely set the direction of the LG's peculiar motion throughout our study volume.

Furthermore our analysis has identified four dynamically interesting CIZA clusters: C1324 in the GA region, C1410 in the Shapley supercluster and the two clusters C1652 and C1638 (Triangulum Australis) which lie behind the GA region at roughly the same distance as Shapley but well in the ZOA.  Lastly, using the LG's peculiar velocity as measured from the CMB anisotropy (corrected for Virgo-centric infall) and the amplitude of the cluster dipole without Virgo included, we find the average value of the $\beta$ parameter to be $\beta = 0.24 \pm 0.02$, in agreement with previous values determined from the XBACs catalogue alone. 
  

On a final cautionary note, we remind the reader that the sparseness of X-ray bright clusters causes them to be noisy tracers of the mass distribution when very small volumes are considered, which inturn causes the dipole zero-point to be ill-constrained.  Therefore one should bear in mind that our conclusions regarding the relative contributions of different distance scales to the final dipole amplitude are valid only in the range we sample well: $40$ to $240 h^{-1}$ Mpc.  That being said, we can think of two methods by which our results can be extended to shallower distances: (i) since truly X-ray selected cluster catalogues such as the BCS have been shown to better sample the nearby cluster distribution, the use of an all-sky, X-ray selected cluster sample would allow a dipole analysis to be performed without the need for weights to compensate for the residual incompleteness in the XBACs sample.  The construction of such a sample will be possible in the near future when the REFLEX catalogue becomes available; its combination with the BCS and CIZA samples will produce the first entirely X-ray selected, all-sky data set.  (ii) Nearby galaxy catalogues can be used to establish a zero-point for the X-ray cluster dipole, from which the relative contributions of different distance scales can truly be determined.  This galaxy dipole normalization has been applied to optical cluster catalogues (Scaramella et al. 1994), but has yet to be implemented on a X-ray selected cluster sample.

\acknowledgments
We greatly thank Brent Tully and Mike Hudson for many useful discussions and contributions, as well as the NASA Graduate Student Research Program for supporting this work.

\clearpage

\end{document}